\documentclass[sn-nature]{sn-jnl}

\usepackage{graphicx}%
\usepackage{amsmath,amssymb,amsfonts}%
\usepackage{amsthm}%
\usepackage{mathrsfs}%
\usepackage[title]{appendix}%
\usepackage{xcolor}%
\usepackage{textcomp}%
\usepackage{manyfoot}%
\usepackage{booktabs}%
\usepackage{algorithm}%
\usepackage{algorithmicx}%
\usepackage{algpseudocode}%
\usepackage{listings}%

\usepackage{hyperref}
\usepackage{float}
\usepackage{placeins}
\usepackage{dcolumn}
\usepackage{bm}
\usepackage{siunitx}
\usepackage{mathtools}
\usepackage[normalem]{ulem}
\usepackage{lipsum}

\usepackage[skip=2pt,font=footnotesize]{caption}
\usepackage{accents}
\newcommand{\via}{{\textit{via}~}}
\newcommand{\ie}{{\textit{i.e.}~}}
\newcommand{\apriori}{{\textit{a priori}~}}
\newcommand{\rmi}[1][]{\mathrm{i}}

\newcommand{\Real}{{\mathrm{Re}}}
\newcommand{\Log}{{\mathrm{Log}}}
\newcommand{\Imag}{{\mathrm{Im}}}

\newcommand{\sio}[1][]{$\mathrm{SiO}_2$~}
\newcommand{\tio}[1][]{$\mathrm{TiO}_2$~}
\newcommand{\ta}[1][]{$\mathrm{Ta}_2\mathrm{O}_5$}
\newcommand{\defeq}{\vcentcolon=}

\usepackage{xcolor}

\usepackage{enumitem}

\theoremstyle{thmstyleone}%
\theoremstyle{thmstyletwo}%

\theoremstyle{thmstylethree}%

\begin{document}

\title[Article Title]{Extracting Complete Resonance Characteristics From the Phase of Physical Signals}


\author*[1]{Isam Ben Soltane}\email{isam.ben-soltane@fresnel.fr}

\author*[1]{Nicolas Bonod}\email{nicolas.bonod@fresnel.fr}

\affil[1]{Aix Marseille Univ, CNRS, Centrale Med, Institut Fresnel, \postcode{13013}, \city{Marseille}, \country{France}}

\abstract{
Resonances are common in wave physics and their full and rigorous characterization is crucial to correctly tailor the response of a system in both time and frequency domains. However, they have been conventionally described by the quality factor, a real-valued number quantifying the sharpness of a single peak in the amplitude spectrum, and associated with a singularity in the complex frequency plane. But the amplitude of a physical signal does not hold all the information on the resonance and it has not been established that even the knowledge of the full distribution of singularities carries this information. Here we derive a dimensionless quality function that fully characterizes resonances from the knowledge of the phase spectrum of the signal. This function is driven by the spectral derivative of the phase. It is equivalent to the imaginary part of the Wigner-Smith time delay but it has the advantage of being valid for arbitrary response functions, including the components of the S-matrix. The spectral derivative of the phase can be calculated numerically from simulations or experimental acquisitions of the phase spectrum. Alternatively, it can be retrieved from the distribution of poles and zeros in the complex frequency plane through an analytic expression, which demonstrates that singularities do not suffice to fully characterize the resonances and that both singularities and zeros must be taken into account to retrieve the quality function. This approach permits to extract all the characteristics of resonances from arbitrary spectral response functions without $\apriori$ knowledge on the physical system.
}




\maketitle

\section{Introduction}
Resonances result from an efficient coupling between an excitation wave and a physical system. They strongly impact the response of the system in both harmonic and time domains and are at the core of many important research fields in wave physics. 
Resonances are at the basis of cavity-enhanced interactions for laser developments cavities~\cite{altug2006ultrafast,kodigala2017lasing,contractor2022scalable}, for cavity electrodynamics~\cite{claudon2010highly,haroche2020cavity,owens2022chiral}, cavity opto-mechanics~\cite{barzanjeh2022optomechanics,vijayan2024cavity}, acoustics~\cite{zalalutdinov2021acoustic}, but also for studying the stability of the response of systems~\cite{dazzo_linear_1983,ben_soltane_derivation_2022,konoplya_quasinormal_2011,van_der_auweraer_discriminating_2004,mihalache_linear_1999}, achieving strong coupling~\cite{frisk2019ultrastrong}, detecting gravitational waves~\cite{greve2022entanglement}, enhancing light-matter interactions~\cite{frisk2019ultrastrong,lalanne2018light,qin_quantum_2024}, or also for tuning the wavefront of light beams~\cite{li2019phase,colom_crossing_2023,lin_universal_2023,malek_multifunctional_2022,barton_wavefront_2021}.

Classically, the strength of the resonance is quantified by the Quality factor ($Q$ factor).
This factor is originally derived from the energy balance within a cavity, which is related to the spectral behaviour of that cavity. As a consequence, the $Q$ factor can be determined from this spectral response as the ratio between the maximum of the response (in amplitude) and the width at mid-height~\cite{green_story_1955,pertersan_measurement_1998,gras2019quasinormal,wu_nanoscale_2021,zambrana-puyalto_quality_2024}. In addition, the introduction of cavity modes often associated with complex frequencies called poles or singularities is common and results in an alternative definition of the $Q$ factor of a resonance associated with a pole $p$ as $\frac{\Real[p]}{2\Imag[p]}$~\cite{neviere_homogeneous_1980,wu_nanoscale_2021}.

These definitions of the $Q$ factor are only equivalent in the rare case of Lorentzian-shaped amplitude curves associated with poles which are isolated from other singularities in the complex $\omega$ plane. However, the response functions of physical systems usually show more complex variations with no apparent symmetry, making it impossible to graphically calculate the $Q$ factor. In addition, the previous definitions assume that the resonant effects are brought about by poles only, disregarding the role of zeros.

Resonances have been associated with the Wigner-Smith time delay in the context of the $S$-matrix scattering formalism. The Wigner-Smith time delay, $\tau_{\omega}$, describes the interaction time of the wave with the scatterer before leaking through a scattering channel in the far field. This concept was first developed in quantum physics to estimate the interaction time between a particle and a potential well~\cite{wigner1955lower,smith1960lifetime}, before being widely used in wave physics and in particular in electromagnetism~\cite{chen_generalization_2021,huang2022wave,giovannelli_physical_2024}. This concept has been associated with the characterization of group delays in multiport systems~\cite{patel2020wigner}, or with the notion of density of states~\cite{davy2015transmission,grabsch2018wigner}. $\tau_{\omega}$ is a real quantity in unitary scattering systems and becomes complex in non-Hermitian systems. The Wigner-Smith time delay $\tau_{\omega}$ depends on the spectral derivative of the phase shift experienced by the wave in the scattering system and can be cast as $\tau_{\omega}=-\frac{i}{M}\frac{d}{d\omega}\Log\left[\mathrm{det}(S)\right]$. Important efforts have been carried out to find the expression of $\tau_{\omega}$ in terms of poles and zeros of the scattering matrix~\cite{chen_generalization_2021,giovannelli_physical_2024}. The most recent advances show how the study of the real and imaginary part of $\tau_{\omega}$ can bring information on poles and zeros of the scattering matrix~\cite{kang2021transmission,chen2021statistics,chen_use_2022}. The link between this complex time delay and resonances is usually achieved by considering modes as equivalent to resonances and poles and zeros. 

Here, we show how the frequency-derivative of the phase of any response function, including the elements of the $S$-matrix, permits to rigorously characterize the resonances of the associated physical system in the whole spectrum. By describing the response function with the Singularity and Zero Factorization (SZF)~\cite{grigoriev_optimization_2013,grigoriev_singular_2013,ben_soltane_multiple-order_2023}, we write the derivative of the phase as a meromorphic function and study the contribution of each complex singularity and zero to its variations, \ie to the shape and the position of the resonances. We derive a quality function which quantifies the resonances and their effect at any frequency $\omega$. This dimensionless function depends on the spectral derivative of the phase, but can also be calculated independently using the poles and zeros extracted from the response function. The accuracy of these two approaches is assessed in several optical cavities and we show that they match very accurately. We analyze the different roles of poles and zeros for characterizing resonances.

\begin{figure}
    \centering
    \includegraphics[width=\textwidth]{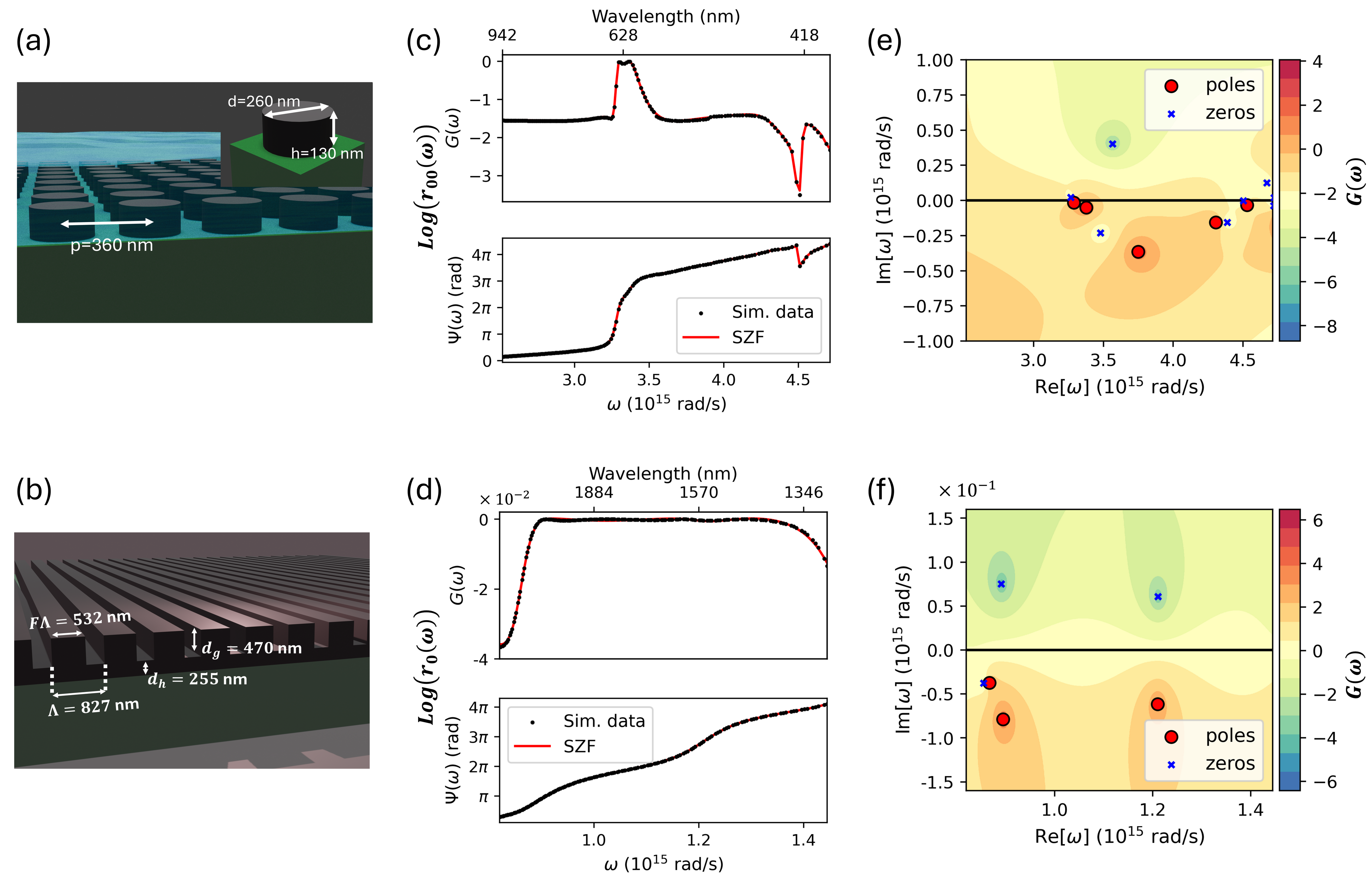}
    \captionsetup{justification=centering}
    \caption{\textbf{Reflection spectrum in amplitude and phase of two optical nanostructures and their associated distribution of zeros and poles in the complex $\omega$-space}. (a) Array of \tio nanodisks of period $p=360$ nm, diameter $d=260$ nm, and height $h=130$ nm, over a substrate of glass, and immersed in water. (b) Grating of Si on a thin layer of Si with thickness $d_h=255$ nm, over a substrate of glass, with a period $\Lambda=827$ nm, a duty cycle $F=64 \%$, and a height $d_g=470$ nm. (c, d) Bode diagrams of the specular reflection coefficients $r_{00}(\omega)$ (a) and $r_{0}(\omega) (b)$ for an illumination at a linearly polarized normal incidence, in the transverse-magnetic polarization. The illumination comes from the glass in the 2D \tio grating, and from the air in the 1D Si grating. (e, f) Log-amplitudes of the reflection coefficients in the complex $\omega$ plane. The associated poles and zeros are indicated by the red and blue dots. They are used to reconstruct the amplitude and phase (SZF, red curves) of the two reflection coefficients in (c,d),through Equation~\eqref{eq:4_2_SZF}.}
    \label{fig:4_1}
\end{figure}

\section{Characterizing resonances with poles and zeros}

Any response function $H(\omega)$ is a complex-valued quantity at real and complex frequencies, thus reading as:
\begin{equation}
    \begin{gathered}
        H(\omega) = |H(\omega)|e^{\rmi \Psi(\omega)} \\
        \Log[H(\omega)] = G(\omega) + \rmi \Psi(\omega)
    \end{gathered}
\label{eq:4_1_complex}
\end{equation}
with $|H(\omega)|$ the amplitude, $G(\omega)$ the log-amplitude, and $\Psi(\omega)$ the phase. $H(\omega)$ possesses complex zeros $z^{(\ell)}$ and poles $p^{(\ell)}$, which can be used to write an exact analytical expression \via the SZF:
\begin{equation}
    H(\omega) = H_0 ~ \frac{\prod_{\ell}(\omega - z^{(\ell)})}{\prod_{\ell}(\omega - p^{(\ell)})} 
\label{eq:4_2_SZF}
\end{equation}
where $H_0$ is a known real-valued constant~\cite{ben_soltane_multiple-order_2023}. Using Equations~\eqref{eq:4_1_complex} and \eqref{eq:4_2_SZF}, the log-amplitude $G(\omega)$ and the phase $\Psi(\omega)$ can be obtained at real frequencies to draw Bode diagrams, or in the complex $\omega$ plane to get complex amplitude and phase maps (or complex Bode diagrams). These plots are drawn in Figure~\ref{fig:4_1} for two optical systems, made of Si and \tio over a glass substrate, and the response function of both systems is their reflection coefficient: (a, c, e) a metasurface composed of an array of nanodisks of \tio in water; (b, d, f) a grating designed to obtain a flat amplitude~\cite{yoon_2015}, associated with a Gires-Tournois effect. The poles and zeros in the complex frequency plane can be retrieved through different means~\cite{baranov2017coherent,sarkar_cauchy_2021,chen_use_2022,ferise_optimal_2023,binkowski2024poles,betz_efficient_2024}. Here they are obtained by fitting the simulated response function $H(\omega)$ with the SZF (red curves) introduced in Equation~\eqref{eq:4_2_SZF}, \via auto-differentiation~\cite{ben_soltane_generalized_2024}, and are shown in the complex amplitude map in Figure~\ref{fig:4_1} (e) and (f). The distributions of log-singularities, \ie poles and zeros, directly set the shape of the spectral responses. The amplitude of the reflection coefficient of the metasurface in Figure~\ref{fig:4_1} (a) exhibits several maxima and minima, related to the alternating poles and zeros close to the real frequency axis in Figure~\ref{fig:4_1} (e). For the Gires-Tournois grating, the spectral range of quasi-constant amplitude is associated with a zero and a pole, which are complex conjugate of each other, and are both isolated from other log-singularities~\cite{mikheeva_2023}.

In both cases, the amplitude spectra do not feature true Lorentzian shapes. In the case of the grating, the presence of resonances cannot even be inferred from the amplitude curve. In both systems, the classical means to characterize resonances using the amplitude cannot be used, and we will now that even the poles do not suffice to characterize the resonance through the ratio between their real and imaginary parts. 

Traces of the resonances must be searched for in the phase plots of the Bode diagrams, in Figure~\ref{fig:4_1} (c, d), where considerable phase variations are observed. Let us stress that the phase of the reflection coefficient of the grating shows a (near) $2\pi$ phase gain over the constant amplitude range~\cite{magnusson_2014,yoon_2015,mikheeva_2023}.

To further investigate the content encoded within the phase, we derive an analytical expression of $\Psi(\omega)$. From Equations~\eqref{eq:4_1_complex}, we have $\Psi(\omega)$ as the imaginary part of the principal value complex logarithm function $\Log$ of $H(\omega)$: 
\begin{equation}
    \Psi(\omega) \equiv \Imag\left[ \Log\left[ H(\omega) \right] \right] \pod{2\pi}
\label{eq:4_3_Psi_formula}
\end{equation}
which yields, with Equation~\eqref{eq:4_2_SZF}, the following expression:
\begin{equation}
    \Psi(\omega) \equiv + \frac{\rmi}{2} \sum_\ell \Log\left[ \frac{\omega-p^{(\ell)}}{\omega-\overline{p^{(\ell)}}} \right] - \frac{\rmi}{2} \sum_\ell \Log\left[ \frac{\omega-z^{(\ell)}}{\omega-\overline{z^{(\ell)}}} \right] \pod{2\pi} 
\label{eq:4_4_Psi}
\end{equation}
$\Psi(\omega)$ appears as a meromorphic function. However, the presence of the $\Log$ function hints at a non-unique definition of the phase, associated with branch cuts in the complex $\omega$ plane~\cite{ablowitz2003}. This is naturally solved for by differentiating Equation~\eqref{eq:4_4_Psi} with respect to the frequency $\omega$:
\begin{equation}
    \partial_\omega[\Psi](\omega) = +\frac{\rmi}{2} \sum_\ell \left[ \frac{1}{\omega - p^{(\ell)}} - \frac{1}{\omega - \overline{p^{(\ell)}}} \right] -\frac{\rmi}{2} \sum_\ell \left[ \frac{1}{\omega - z^{(\ell)}} - \frac{1}{\omega - \overline{z^{(\ell)}}} \right]
\label{eq:4_5_dPsi_dw}
\end{equation}
Equation~\eqref{eq:4_5_dPsi_dw} highlights the crucial role of the zeros $z^{(\ell)}$, symmetric to that of the poles $p^{(\ell)}$, in the calculation of $\partial_\omega[\Psi](\omega)$ (and the phase $\Psi(\omega)$ by extension). Equation~\eqref{eq:4_5_dPsi_dw} is a singularity expansion~\cite{baum_singularity_1991,ben_soltane_multiple-order_2023,sol2023reflectionless,jiang2024coherent} of $\partial_\omega[\Psi](\omega)$, which implies that the poles and zeros fully characterizes $\partial_\omega[\Psi](\omega)$~\cite{chen_generalization_2021,erb_control_2023,giovannelli_physical_2024}. An important consequence is that the zeros are as important as poles in the description of resonances, \ie the poles do not suffice in characterizing resonances.

\begin{figure}
    \centering
    \includegraphics[width=.7\textwidth]{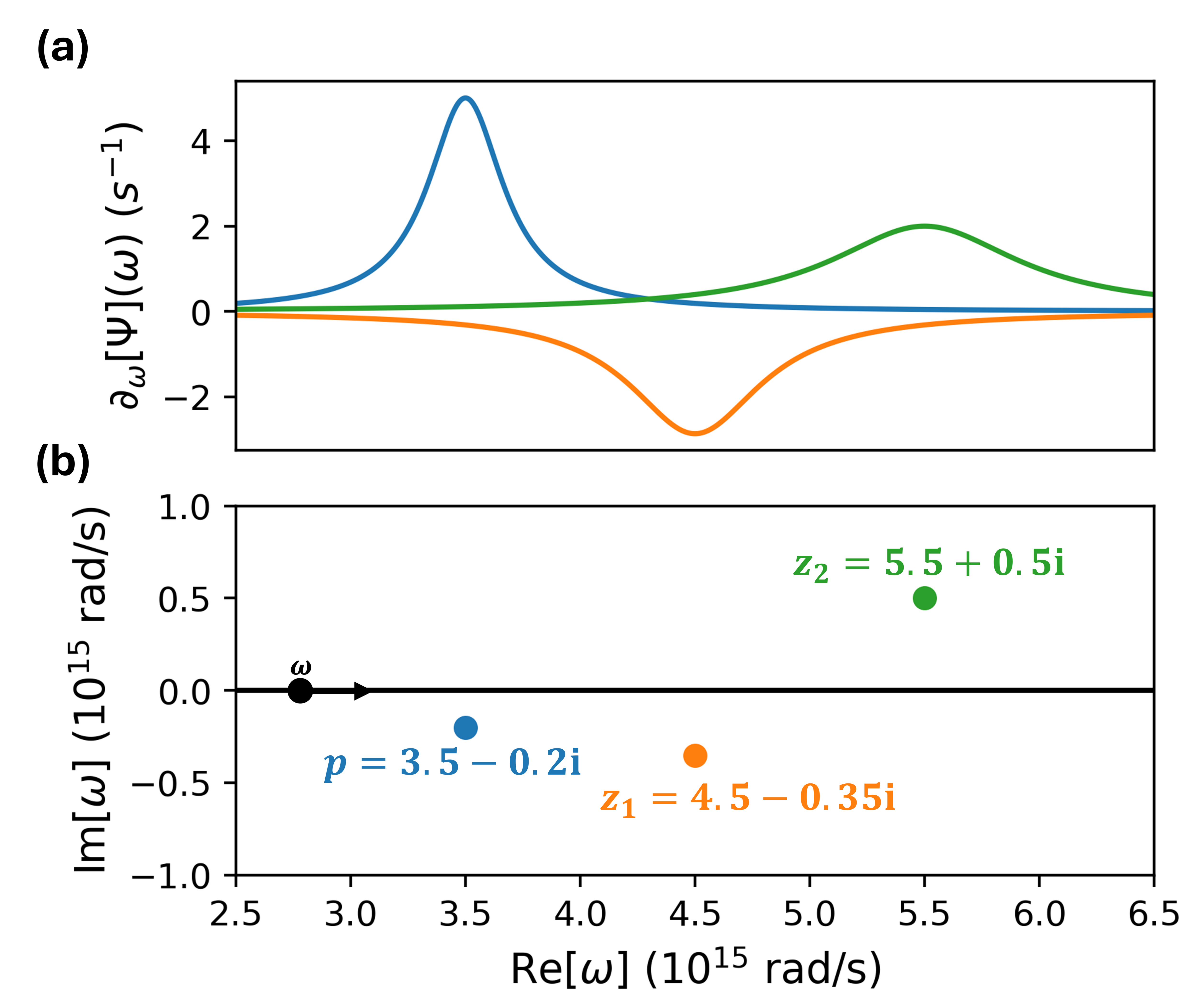}
    \captionsetup{justification=centering}
    \caption{\textbf{Contribution of poles and zeros to the spectral derivative of the phase: poles located in the lower half $\omega$-space always have a positive contribution while zeros can have positive or negative contribution if located in the upper or lower half $\omega$-space respectively.}. (a) Differentiated-phase term $D_s(\omega)$ for a pole $s=p$ (blue), and two zeros $s=z_1$ (orange) and $s=z_2$ (green). (b) The positions of the log-singularities are shown in the complex $\omega$-plane. The sharpness and peak-value of each resonant term are directly related to the imaginary part of the log-singularity, \ie to their distance to the real frequency axis (black horizontal line).}
    \label{fig:4_2}
\end{figure}

Let us now investigate each term $D_s(\omega)$ in the expression of $\partial[\Psi](\omega)$ in Equation~\eqref{eq:4_5_dPsi_dw}:
\begin{equation}
    D_s(\omega) = \frac{\rmi}{2} \left[ \frac{1}{\omega - s} - \frac{1}{\omega - \overline{s}} \right] = -\frac{\Imag[s]}{\left|\omega - s\right|^2}
\end{equation}
where $s=p^{(\ell)}$ for the terms involving a pole, and $s=\overline{z^{(\ell)}}$ for those involving a zero. $D_s(\omega)$ is a real-valued function with a perfect Lorentzian shape when evaluated at real frequencies, as shown in Figure~\ref{fig:4_2}, where we calculate this function for one pole $s=p$, and two zeros $s=z_1$ or $s=z_2$. It reaches its extremum value $-\frac{1}{\Imag[s]}$ at $\omega=\Real[s]$. That maximum (or minimum) value is divided by two as $\omega$ is at a distance $|\Imag[s]|$ of the resonance frequency $\Real[s]$.

By integrating $D_s(\omega)$ over the full real $\omega$ axis, from $-\infty$ to $+\infty$, a $\pm\pi$ phase-shift is obtained, for any log-singularity, regardless of its type (zero or pole) and its position~\cite{grigoriev_optimization_2013,grigoriev_singular_2013,colom_crossing_2023}. We show that a phase-variation $\Delta\Psi_B=\pm \pi/2$ is obtained by integrating over  the smaller $\left[- |\Imag[s]|, + |\Imag[s]| \right]$ spectral range, centered around the real part of the log-singualrity $\Real[s]$, which we refer to as the bandwidth of the log-singularity:
\begin{equation}
    \begin{aligned}
        \Delta\Psi_B &= \int_{\Imag[s]}^{-\Imag[s]} D_s(\Real[s] + \omega) = \pm \frac{\rmi}{2} \Log\left(\frac{\Imag[s](1-\rmi)(-1-\rmi}{\Imag[s](1+\rmi)(-1+\rmi}\right) = \pm \frac{\pi}{2}
    \end{aligned}
\end{equation}

This phase variation is a phase gain if $\Imag[s]<0$, and a phase loss if  $\Imag[s]>0$. Poles are always assumed to be stable and have thus a negative imaginary part. As a consequence, a phase gain is observed. Zeros can have either a positive or negative imaginary part, \ie $\Imag[s]=-\Imag[z^{(\ell)}]$ is positive if the zero is in the upper part of the complex frequency plane, yielding a $\pi/2$ phase gain over the aforementioned range. Otherwise, a phase loss is observed. The resonance frequencies are thus identified as peaks in the phase-derivative curve, while the amplitude and width of these peaks are directly related to the imaginary part of the pole or zero. Let us point out that a phase loss is always brought about by the presence of zeros in the lower part of the complex $\omega$ plane. It is however impossible to discriminate between a pole and a non-reversible zero, \ie a zero with a positive imaginary part, when a phase gain is observed, in which case the amplitude is required to identify a pole or zero, as described in Section I of the Supporting Information (SI).

\section{Quality function of an arbitrary response function}

At high frequencies, fields oscillate faster within the systems, allowing them to interact over smaller time periods. Therefore, their amplitude varies faster than at low frequencies. Consequently, the poles and zeros tend to move away from the real $\omega$ axis as we move into higher spectral domains.

Each term $D_s(\omega)$ quantifies the proximity of the frequency of interest $\omega$ to the log-singularity $s$. It is thus often lower-valued at high frequencies than at low frequencies. As a response to that effect, we introduce the quality function of the response function, defined as the non-dimensional phase variation function:
\begin{equation}
    \eta(\omega) \defeq \frac{\omega}{2}\partial_\omega[\Psi](\omega)
\label{eq:4_7}
\end{equation}

\begin{figure}
    \centering
    \includegraphics[width=\textwidth]{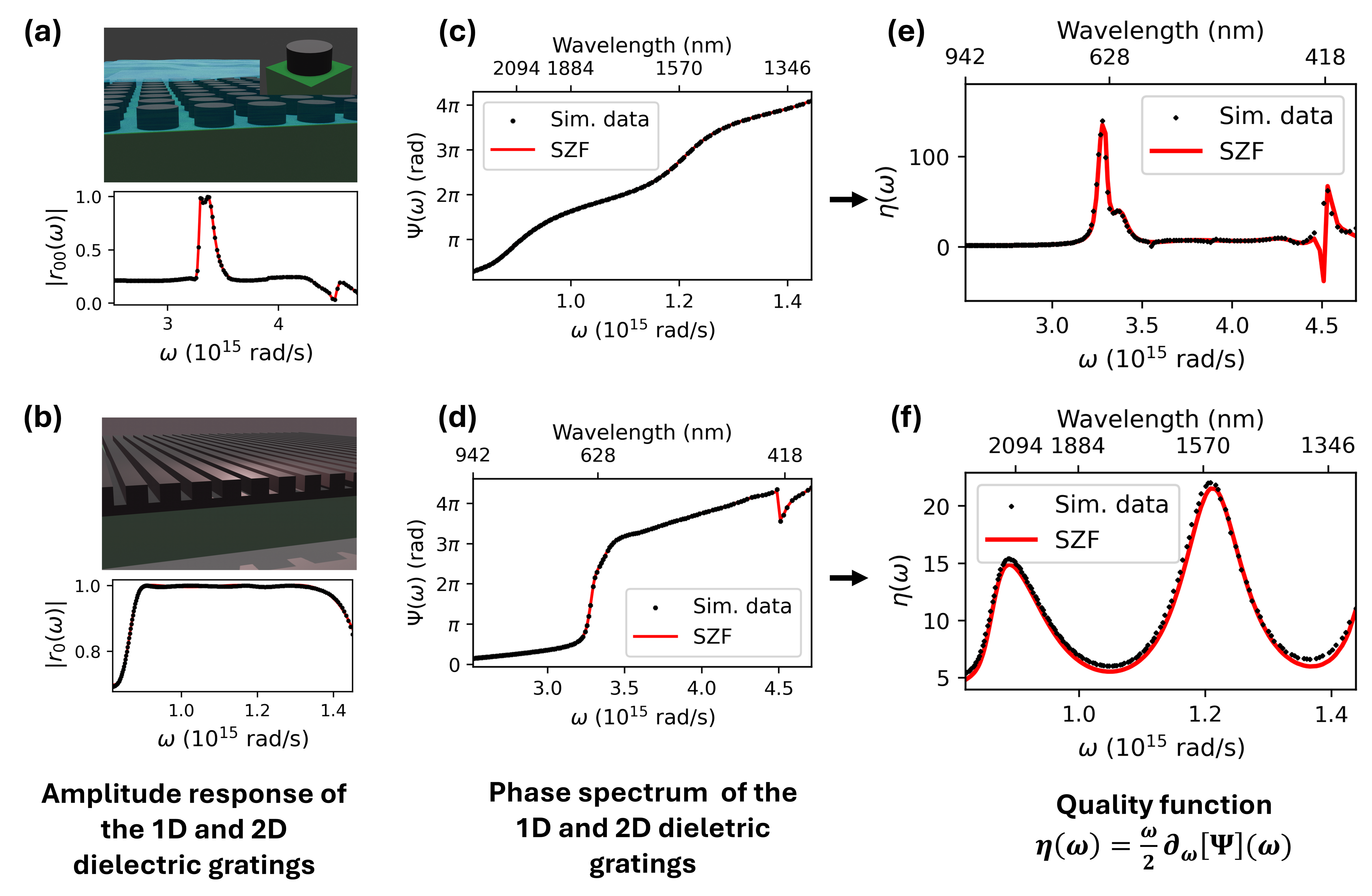}
    \captionsetup{justification=centering}
    \caption{\textbf{Extraction of the quality function $\eta(\omega)$ from the phase spectrum of the optical system from the numerical calculation of the spectral derivative of the phase (black dots) or with the poles and zeros (red line).}
    (a,b) Amplitude response of the two systems introduced in Figure~\ref{fig:4_1}, \ie the array of nanodisks of \tio in (a), and the Gires-Tournois grating in (b). (c,d) Phase spectrum associated to the two optical systems. Simulated data in black dots are compared with the reconstruction through poles and zeros in the SZF expression from Equation~\ref{eq:4_2_SZF}. (e,f) Quality function $\eta(\omega)$ calculated for these two systems. The quality function $\eta(\omega)$ is obtained directly from the experimental data with Equation~\ref{eq:4_7} using a first-order divided difference. It is compared to the meromorphic singularity expansion in Equation~\eqref{eq:4_8} using the poles and zeros in the SZF expression~\ref{eq:4_2_SZF}.}
    \label{fig:4_3}
\end{figure}

The calculation of the quality function $\eta(\omega)$ using Equation~\eqref{eq:4_7} with experimental data is straightforward. The frequency-derivative of $\Psi(\omega)$ is numerically approximated \via a first-order divided difference. The results are displayed in Figure~\ref{fig:4_3} (e,f) for the metasurface and the grating introduced in Figure~\ref{fig:4_1}. Each extremum in the curve of $\eta(\omega)$ corresponds to a resonance. In the case of the metasurface, this approach provides a highly accurate localization of the resonances, which cannot be done through classically by considering the amplitude or phase curves in the Bode diagram. The relevancy of the quality function is even more striking in the case of the Gires-Tournois grating: Theresonances are clearly highlighted and quantified by the quality function $\eta(\omega)$, while they are invisible in the Bode diagram. The quality function unveils the resonances of any arbitrary spectral response function. Let us stress that $\eta(\omega)$ reveals the information encoded in the phase of the spectral response without any \apriori knowledge on the physical system itself.

Let us now expand the quality function in terms of the poles and zeros of $H(\omega)$. Since $\partial\Psi(\omega)$ is a meromorphic function, $\eta(\omega)$ is also a meromorphic function, which means that it is also fully characterized by the poles and zeros of $H(\omega)$:
\begin{equation}
    \eta(\omega) = +\frac{\rmi}{4} \sum_\ell \left[ \frac{\omega}{\omega - p^{(\ell)}} - \frac{\omega}{\omega - \overline{p^{(\ell)}}} \right] - \frac{\rmi}{4} \sum_\ell \left[ \frac{\omega}{\omega - z^{(\ell)}} - \frac{\omega}{\omega - \overline{z^{(\ell)}}} \right]
\label{eq:4_8}
\end{equation}
We show in Figure~\ref{fig:4_3} (e,f) that this description \via the poles and zeros matches the function calculated directly with experimental data.

Similarly to the previous study of the function $\partial_\omega[\Psi](\omega)$, we consider the role of each log-singularity $s$ to the function $\eta(\omega)$, \via the terms $\eta_s(\omega)\defeq \frac{\omega}{2} D_s(\omega)$: 
\begin{equation}
    \eta_s(\omega) = -\frac{\omega\Imag[s]}{(\sqrt{2}|\omega-s|)^2}
\label{eq:4_11}
\end{equation}
Each term $\eta_s(\omega)$ has one extremum at the resonance frequency $\omega=\Real[s]$:
\begin{equation}
    \eta_s(\Real[s]) = -\frac{\Real[s]}{2\Imag[s]}
\label{eq:4_10_Qs}
\end{equation}
When the log-singularity $s$ is a pole, Equation~\eqref{eq:4_10_Qs} identifies with the commonly used $Q$ factor $Q_{p^{(\ell)}}$:
\begin{equation}
    \eta_{p^{(\ell)}}(\Real[p^{(\ell)}]) =  Q_{p^{(\ell)}} >0
\end{equation}
In the case of a zero, we obtain a definition of the classical $Q$ factor $Q_{z^{(\ell)}}$ of the zero $z^{(\ell)}$, which can be negative if the zero is located in the lower part of the complex $\omega$ plane. The classical $Q$ factor $Q_s$ of an isolated resonance therefore corresponds to a particular case of $\eta_s(\omega)$, for $\omega=\Real[s]$.

\section{Phase variation analysis with poles and zeros}

Equation~\eqref{eq:4_11} shows $\eta(\omega)$ as a non-dimensional function, corresponding to the ratio of two quantities with the same unit, \ie $-\omega\Imag[s]$ and $(\sqrt{2}|\omega-s|)^2$. These quantities are the areas of two surfaces in the complex frequency plane, as shown in Figure~(S1) in the SI, each associated with a different effect.

\begin{figure}[H]
    \centering
    \includegraphics[width=\textwidth]{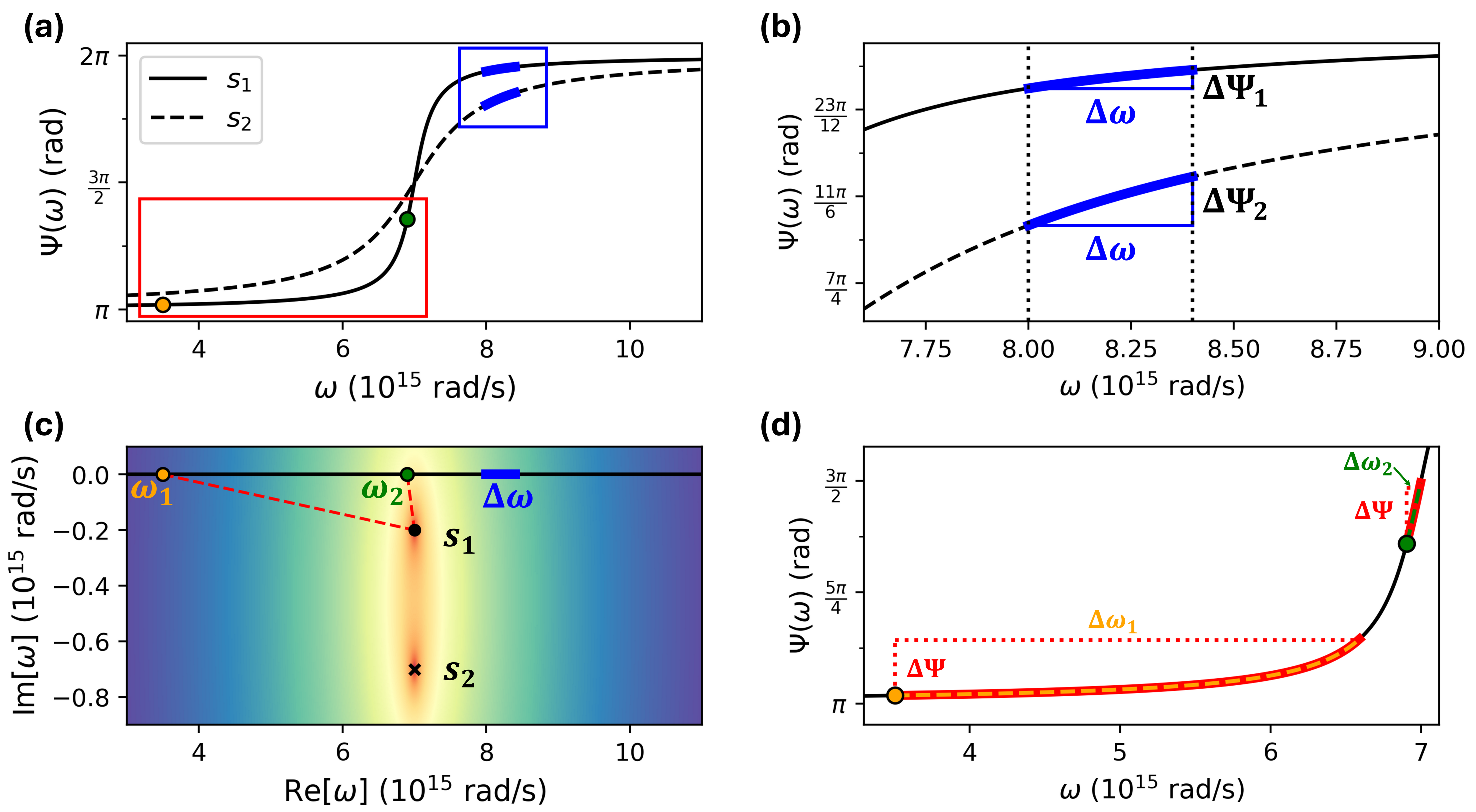}
    \captionsetup{justification=centering}
    \caption{\textbf{Graphical interpretation }. (a) Phase $\Psi(\omega)$ of the function $H(\omega)=\frac{1}{\omega - s}$ for two log-singularities $s_1$ and $s_2$. The yellow and green dots on the curve of $s_1$ are associated with the real frequencies $\omega_1$ and $\omega_2$. The blue lines correspond to a spectral range of width $\Delta\omega$. (b) Position of the log-singularities (black dot and black cross), of $\omega_1$ and $\omega_2$, and of $\Delta\omega$ in the complex frequency plane. The distance from $\omega_2$ to $s_1$ is smaller than that from $\omega_1$ to $s_1$ (dashed red lines). (c) Phase functions of both $s_1$ (full black curve) and $s_2$ (dashed black curve). For a fixed frequency variation $\Delta\omega$, a phase-shift $\Delta\Psi_1$ is obtained for the log-singularity $s_1$, and $\Delta\Psi_2$ for $s_2$, with $\Delta\Psi_2 \ge \Delta\Psi_1$. (d) Phase function associated with $s_1$ only. To obtain a fixed phase variation $\Delta\Psi=\frac{\pi}{8}$, a frequency shift $\Delta\omega_1$ is required as we start from $\omega_1$ (yellow dot), greater than the frequency shift $\Delta\omega_2$ required when we start from $\omega_2$ (green dot).}
    \label{fig:4_4}
\end{figure}

(i) The numerator $-\omega\Imag[s]$ is associated with the distance from the log-singularity $s$ to the real frequency axis. It quantifies the information or phase-variations that can be retrieved over the real frequency axis. To dissociate its effect from that of the denominator $(\sqrt{2}|\omega-s|)^2$, we set $\omega$ far enough from the log-singularity $s$ for the term $|\omega - s|$ to be almost constant. The effect of the numerator is quantified by the phase variations $\Delta\Psi_1$ and $\Delta\Psi_2$ in Figure~\ref{fig:4_4} (b), associated with the log-singularities $s_1$ and $s_2$ in the complex $\omega$ plane (see Figure~\ref{fig:4_4} (b)), each corresponding to a resonance, with $s_1$ being closer to the real frequency axis than $s_2$. These phase-shifts are induced by a fixed frequency shift $\Delta\omega$ at an arbitrary frequency $\omega$. We see in Figure\ref{fig:4_4} (b) that $\Delta\Psi_1 \le \Delta\Psi_2$ which shows that the further the log-singularity is from the real $\omega$ axis, the bigger the associated phase-shift $\Delta\Psi$. This explains why the quantity in the numerator grows as $|\Imag[s]|$ increases.

(ii) The denominator is related to the distance $|\omega - s|$. A log-singularity $s$ introduces a $\pm \pi$ phase-shift as $\omega$ goes from $-\infty$ to $+\infty$. In order to quantify how easily that potential phase-shift can be exploited at a given frequency $\omega$, we look, in Figure~\ref{fig:4_4} (d) at the frequency variations $\Delta\omega_1$ and $\Delta\omega_2$ required to obtain a fixed phase-shift $\Delta\Psi$ (red curves), starting from the two real frequencies $\omega_1$ (yellow dot) and $\omega_2$ (green dot). The distance from $\omega_2$ to the log-singularity $s_1$ is greater than that from $\omega_2$ to $s_2$, as shown in Figure~\ref{fig:4_4} (c), and $\Delta\omega_1 \geq \Delta\omega_2$ in Figure~\ref{fig:4_4} (c). It turns out that the further $\omega$ is from the log-singularity $s$, the bigger $\Delta\omega$ is. The quantity in the denominator thus becomes smaller when $\omega$ is close to $\Real[s]$.

The real-valued quality function $\eta(\omega)$ thus quantifies how much of the phase variation resulting from the presence of each log-singularity can be exploited at any frequency.

\section{Conclusion}

To conclude, resonances are fully characterized by the frequency derivative of the phase $\partial_\omega[\Psi](\omega)$ of a response function $H(\omega)$. The phase of the physical signal carries all the information regarding the resonances. 
This information can be simply decoded by calculating the quality function, a function obtained by differentiating the spectral phase function, and simply given by $\eta(\omega) \defeq \frac{\omega}{2}\partial_\omega[\Psi](\omega)$. This function unveils resonances from arbitrary continuous spectral phase function. The maximal strength of resonances is exactly given by the extrema of this quality function. It is a meromorphic function that can be expanded in terms of poles and zeros of the response function $H(\omega)$, providing an analytical continuation into the complex $\omega$ plane. 
This expansion demonstrates that poles alone are not enough to describe resonances and that zeros are also required in order to characterize them. Zeros can have either positive or negative contributions to the quality function depending on their position in the lower (negative contribution) or upper (positive contribution) part of the complex frequency plane; poles being always located in the lower half-plane, they systematically bring about a positive contribution. This shows that the commonly admitted $\pi$ phase-shift induced by a resonance or singularity, both being classically considered as equivalent, should be replaced by a more general analysis of the phase variation brought about by both poles and zeros. Far from being restricted to optical devices, this approach is general and can be applied to any linear system in wave physics. It shows that all resonance characteristics can be extracted from the phase spectrum of a signal without any \apriori knowledge regarding the physical system.

\backmatter

\bmhead{Supplementary information}
The Supporting Information is available from XXXXXX or from the authors.

\bmhead{Acknowledgements}
This work was funded by the French National Research Agency ANR Project DILEMMA (ANR-20-CE09-0027).

\end{document}


\title[Article Title]{Supporting information for the manuscript entitled ``Extracting Resonance Characteristics from the Phase of Physical Signals''}

\author*[1]{\fnm{Isam} \sur{Ben Soltane}}\email{isam.ben-soltane@fresnel.fr}

\author*[1]{\fnm{Nicolas} \sur{Bonod}}\email{nicolas.bonod@fresnel.fr}

\affil[1]{Aix Marseille Univ, CNRS, Centrale Marseille, Institut Fresnel, \postcode{13013}, \city{Marseille}, \country{France}}

\maketitle

\section{Singularity expansion of the phase and log-amplitude derivatives}

\noindent
The response function $H(\omega)$ of a physical system is a complex-valued quantity at real and complex frequencies, which reads as:
\begin{equation}
    \begin{gathered}
        H(\omega) = |H(\omega)|e^{\rmi \Psi(\omega)} \\
        \Log[H(\omega)] = G(\omega) + \rmi \Psi(\omega)
    \end{gathered}
\label{eq:S_1_complex}
\end{equation}
with $|H(\omega)|$ the amplitude, $G(\omega)$ the log-amplitude, and $\Psi(\omega)$ the phase. $H(\omega)$ possesses complex zeros $z^{(\ell)}$ and poles $p^{(\ell)}$, which can be used to write an exact analytical expression \via the singularity and zero factorization:
\begin{equation}
    H(\omega) = H_0 ~ \frac{\prod_{\ell}(\omega - z^{(\ell)})}{\prod_{\ell}(\omega - p^{(\ell)})} 
\label{eq:S_2_SZF}
\end{equation}
where $H_0$ is a known real-valued constant. From Equations~\eqref{eq:S_1_complex} and \eqref{eq:S_2_SZF}, we see that $G(\omega)$ and $\Psi(\omega)$ are characterized by the log-singularities (singularities of the complex logarithm of $H(\omega)$) zeros $z^{(\ell)}$ and poles $p^{(\ell)}$.  From Equations~\eqref{eq:S_1_complex}, we have $G(\omega)$ and $\Psi(\omega)$ as the real and imaginary parts of the principal value complex logarithm function $\Log$ of $H(\omega)$: 
\begin{align}
    G(\omega) &= \Real\left[ \Log\left[ H(\omega) \right] \right] \label{eq:S_3_G_formula}\\
    \Psi(\omega) &\equiv \Imag\left[ \Log\left[ H(\omega) \right] \right] \pod{2\pi} \label{eq:S_4_Psi_formula}
\end{align}
which yields, with Equation~\eqref{eq:S_2_SZF}, the following expressions:
\begin{align}
    &\begin{aligned}
        G(\omega) = &+ \frac{1}{2} \sum_\ell \Log\left[ \left(\omega-z^{(\ell)}\right)\left(\omega-\overline{z^{(\ell)}}\right) \right]\\
        &- \frac{1}{2} \sum_\ell \Log\left[ \left(\omega-p^{(\ell)}\right)\left(\omega-\overline{p^{(\ell)}}\right) \right]
    \end{aligned}\label{eq:S_5_G}\\
    &\begin{aligned}
        \Psi(\omega) \equiv &+ \frac{\rmi}{2} \sum_\ell \Log\left[ \frac{\omega-p^{(\ell)}}{\omega-\overline{p^{(\ell)}}} \right] \\
        &- \frac{\rmi}{2} \sum_\ell \Log\left[ \frac{\omega-z^{(\ell)}}{\omega-\overline{z^{(\ell)}}} \right] \pod{2\pi} 
    \end{aligned}\label{eq:S_6_Psi}
\end{align}
$G(\omega)$ and $\Psi(\omega)$ are meromorphic functions, but are not uniquely defined due to the presence of the $\Log$ function. However, it turns out that their frequency-derivatives can be explicitly and uniquely written as singularity expansions using the log-singularities of the response function:
\begin{align}
    &\begin{aligned}
        \partial_\omega[G](\omega) = &+\frac{1}{2} \sum_\ell \left[ \frac{1}{\omega - z^{(\ell)}} + \frac{1}{\omega - \overline{z^{(\ell)}}} \right] \\
        &-\frac{1}{2} \sum_\ell \left[ \frac{1}{\omega - p^{(\ell)}} + \frac{1}{\omega - \overline{p^{(\ell)}}} \right]
    \end{aligned} \label{eq:S_7_dG_dw} \\
    &\begin{aligned}
        \partial_\omega[\Psi](\omega) = &+\frac{\rmi}{2} \sum_\ell \left[ \frac{1}{\omega - p^{(\ell)}} - \frac{1}{\omega - \overline{p^{(\ell)}}} \right] \\
        &-\frac{\rmi}{2} \sum_\ell \left[ \frac{1}{\omega - z^{(\ell)}} - \frac{1}{\omega - \overline{z^{(\ell)}}} \right]
    \end{aligned} \label{eq:S_8_dPsi_dw}
\end{align}
In Equations~\eqref{eq:S_7_dG_dw} and \eqref{eq:S_8_dPsi_dw}, each term involving a pole or a zero and their complex conjugate describes the effect of a log-singularity on the variations of the amplitude and phase functions. These variations depend on the distribution or the positions of the poles and zeros. Since the poles are located in the lower part of the complex frequency plane, they always have a positive contribution for both the phase and the log-amplitude. The zeros, on the other hand are either in the upper or lower part of the complex $\omega$ plane. In the former case, they have a positive contribution to the phase and amplitude, just like the poles. In the latter case, when their imaginary part is negative, they contribute negatively to the phase and log-amplitude and have thus a role which is anti-symmetric to that of the poles, as indicated by the change of sign between the sums over the index $\ell$.\\

\noindent
This anti-symmetry of the roles, along with the freedom in the placement of zeros in the complex plane, are the reasons why it is generally not possible to obtain all the information regarding a system or its response through only the phase or the amplitude. In the case of the derivative of the log-amplitude $\partial_\omega[G](\omega)$, a pole can never be mistaken for a zero owed to the anti-symmetry. However, because of the "$+$" sign in within the brackets in Equations~\eqref{eq:S_7_dG_dw} and \eqref{eq:S_8_dPsi_dw}, it is not possible to discriminate between a zero $z^{(\ell)}$ and its complex conjugate $\overline{z^{(\ell)}}$ since $z^{(\ell)}$ can either a positive or a negative imaginary part. For the derivative of the phase $\partial_\omega[\Psi](\omega)$, the opposite is observed. The "$-$" sign within the brackets makes it impossible to mistake a pole or a zero for their complex conjugate. However, it is not possible to discriminate between a pole and a zero with a positive imaginary part. The information contained within the phase and the log-amplitude are thus complementary, and both functions are required to fully characterize $H(\omega)$.

\section{Geometrical interpretation of the quality function}

\begin{figure}
    \centering
    \includegraphics[width=\textwidth]{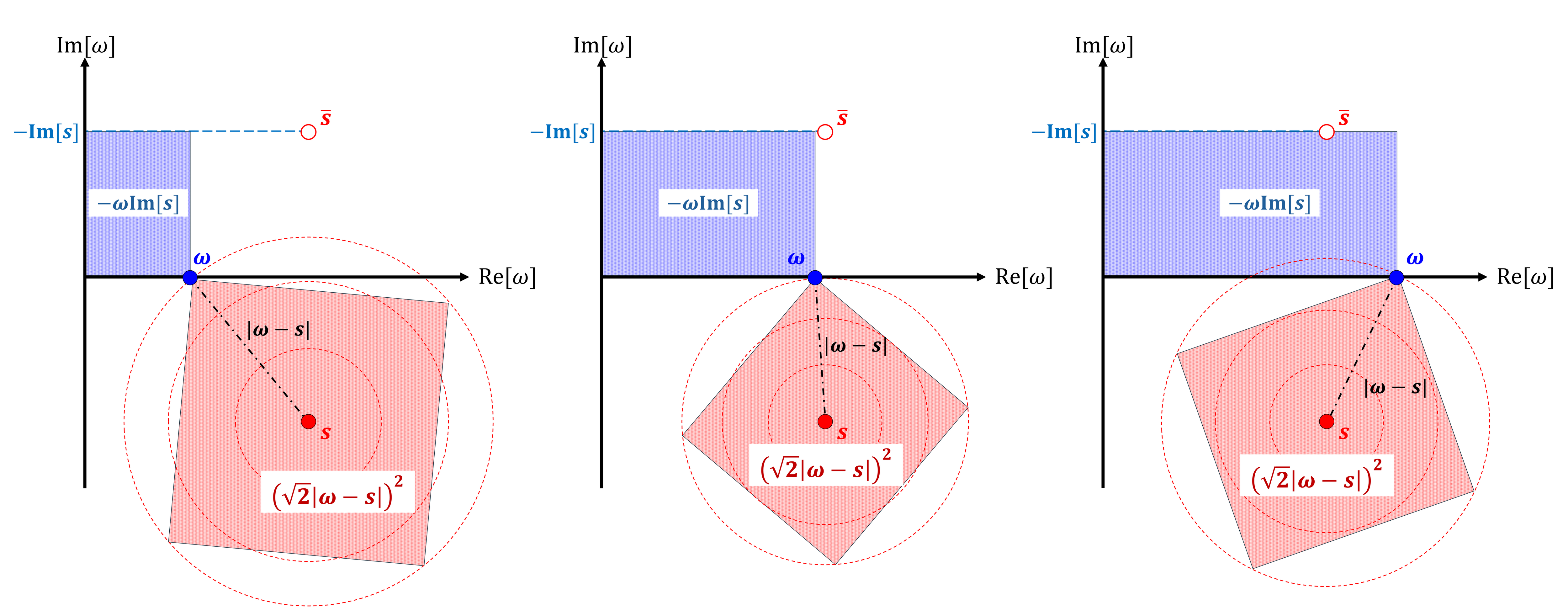}
    \captionsetup{justification=centering}
    \caption{Geometrical interpretation of the quality function term $\eta_s(\omega)$ in Equation~\eqref{eq:S_11}, for a frequency $\omega$ and an arbitrary log-singularity $s$. The numerator $-\omega\Imag[s]$ is the area of the blue rectangle, while the denominator $(\sqrt{2}|\omega - s|)^2$ is the area of the red rectangle in the lower part of the complex $\omega$ plane.}
    \label{fig:4_S1}
\end{figure}

We introduce the quality function $\eta(\omega)$ as:
\begin{equation}
    \eta(\omega) \defeq \frac{\omega}{2} \partial_\omega[\Psi](\omega)
\label{eq:S_9}
\end{equation}
The quality function quantifies the resonances, \ie the variations of the phase, at any frequency $\omega$. Using Equation~\eqref{eq:S_8_dPsi_dw}, we can describe the behaviour of $\eta(\omega)$ with a singularity expansion involving the poles and zeros of $H(\omega)$:
\begin{equation}
    \begin{aligned}
        \eta(\omega) = &+\frac{\rmi}{4} \sum_\ell \left[ \frac{\omega}{\omega - p^{(\ell)}} - \frac{\omega}{\omega - \overline{p^{(\ell)}}} \right] \\
        &-\frac{\rmi}{4} \sum_\ell \left[ \frac{\omega}{\omega - z^{(\ell)}} - \frac{\omega}{\omega - \overline{z^{(\ell)}}} \right]
    \end{aligned}\label{eq:S_10}
\end{equation}
Subsequently, the contribution of each pole or zero is obtained by studying the independent terms $\eta_s(\omega)$ defined as: 
\begin{equation}
    \begin{aligned}
        \eta_s(\omega) &\defeq \left[ \frac{\omega}{\omega - s} - \frac{\omega}{\omega - s} \right] \\
        &= -\frac{\omega\Imag[s]}{(\sqrt{2}|\omega-s|)^2}
    \end{aligned}
\label{eq:S_11}
\end{equation}
where $s$ is a log-singularity, \ie either a pole $p^{(\ell)}$ or a zero $z^{(\ell)}$. A geometrical interpretation of each term of the the quality function is obtained by noticing that the two quantities appearing in the ratio of Equation~\eqref{eq:S_11} are areas of two the surfaces shown in Figure~\ref{fig:4_S1} as a blue rectangle and a red square.\\

\noindent
The blue rectangle grows with the distance $-\Imag[s]$ from the log-singularity $s$ to the real $\omega$ axis, and with the frequency $\omega$ of interest. The red rectangle is proportional to the square of the distance $|\omega-s|$ from the studied real frequency $\omega$ to the log-singularity $s$. It reaches its minimum value at $\omega=\Real[s]$, when the frequency is right above or below the log-singularity.